\documentclass[12pt]{iopart}

\usepackage{iopams}  
\usepackage{graphicx}  

\newcommand{\Npart}{\mbox{$N_{\mathrm{part}}$}}
\newcommand{\Ncoll}{\mbox{$N_{\mathrm{coll}}$}}

\newcommand{\PbPb}{\mbox{Pb+Pb}}

\newcommand{\Rcp}{\mbox{$R_{\mathrm{CP}}$}}
\newcommand{\pT}{\mbox{$p_{\mathrm{T}}$}}

\begin{document}

\title{Centrality dependence of charged particle spectra and $\Rcp$ in \PbPb\ 
collisions at $\sqrt{s_{NN} }=2.76$~TeV with the ATLAS detector at the LHC}

\author{Alexander Milov, for the ATLAS Collaboration}

\address{Department of Particle Physics and Astrophysics,\\
Weizmann Institute of Science, Rehovot 76100, Israel}
\ead{alexander.milov@weizmann.ac.il}
\begin{abstract}
The ATLAS experiment at the LHC measures the charged particle spectra and 
the nuclear modification factor in \PbPb\ collisions at the 
$\sqrt{s_{NN} }=2.76$~TeV in a transverse momentum range up to 30~GeV 
and a pseudorapidity range up to $|\eta|<2.5$. The measurement reveals 
the strong suppression of charged hadron production in the most central 
collisions at a \pT\ of about 7~GeV. A suppression of more than a factor of 
2 is also measured at the upper edge of the analyzed \pT\ range. The 
suppression does not show any strong $\eta$ dependence.
\end{abstract}


\section{Introduction}
High \pT\ hadrons are generally regarded to be produced through jet fragmentation, 
which has been successfully modeled using perturbative QCD. The energy loss of hard 
scattered partons traversing the hot and dense medium translates into a reduction in the 
yield of hadrons, measured as a ratio of yields per nucleon-nucleon interactions in 
head-on heavy ion collisions to those measured in p+p collisions~\cite{Vitev, Eskola}. 
Even without equivalent proton-proton data the ratio of yields can be studied as a function 
of centrality in the heavy ion collisions.  In this case the \Rcp\ is defined as the ratio of yields 
measured in central collisions to the yield measured in peripheral collisions, both scaled by the 
corresponding numbers of binary nucleon-nucleon collisions~\cite{atlas_note}.
Study of the charged particle spectra at high \pT\ provides an independent method 
to understand the jet quenching independent of any particular jet reconstruction algorithm.

\section{Analysis}
\label{sec:dataset}
This analysis uses $4.36\times10^{7}$ minimum bias events ($L_{int}\approx7\mu\mathrm{b}^{-1}$) taken during the 2010 
LHC lead-lead run and satisfying run and event cleaning selections. The event centrality in data is estimated 
using the total transverse energy measured by the ATLAS Forward Calorimeter, as described in~\cite{Mult}.
The correspondence between the cross section measured in the data after all trigger and event selection 
requirements and the cross section estimated using the Glauber model~\cite{CentReview} was found to be 100$\pm$2\%
Using this number, the data is binned in 5\% and 10\% centrality bins and the mean 
number of participants and collisions, \Ncoll, for each centrality bin is estimated 
with the same Glauber calculation.

The Monte Carlo studies are based on the HIJING~\cite{Hijing} and PYTHIA~\cite{Pythia} event generators.
Two samples are used, one containing $2.5\times10^{5}$ HIJING minimum bias events and one with $1.5\times10^{6}$ PYTHIA jet events 
simulated at $\sqrt{s_{NN}}=2.76$ TeV overlaid with minimum-bias HIJING events. 
This sample is used to study track reconstruction efficiency at high \pT. 
Centrality in HIJING events 
is determined to match the measured occupancy in the first 
layer of the ATLAS Pixel detector for each centrality bin.

Charged particle tracks are reconstructed using the ATLAS Inner Detector immersed in 
2~T field of a superconducting solenoid magnet. Tracks are measured in the pseudorapidity region 
$|\eta|<2.5$ 
and over full azimuth above the \pT\ cut of 500~MeV. A charged particle passing through the  
Inner Detector region typically traverses three layers of silicon pixel detectors (Pixel),
and four double-sided silicon strip modules of the semiconductor tracker (SCT)~\cite{ATLASdet}.
To improve the purity of the ID track reconstruction in the dense environment of heavy ion
collisions, track quality requirements are more stringent than those used for 
p+p collisions~\cite{ATLASminb}. We required at least two hits on a track in the Pixel detector 
and at least 8 hits in the SCT with the condition that there are no missing hits in either detector. 
Somewhat more relaxed cuts were also studied to work out the systematic uncertainties. 
This analysis does not make use of information from the Transition Radiation Tracker.

The contribution from combinatorial fake tracks resulting from spurious hit combinations and 
from secondaries is further reduced by requiring the significance of the longitudinal and 
transverse impact parameters of each track to be less than 3. The significance is the track 
impact parameter $d_0$ and $z_0 \sin (\theta)$ divided by the track fit errors arising 
from the reconstruction, which accounts for the \pT\ and $\eta$ dependence of the tracking performance.

The track reconstruction efficiency was evaluated using Monte Carlo samples.
The efficiencies are found to systematically decrease with $|\eta|$ from 0.7 to 0.5 in the 
range of measurement and also show \pT\ dependence below 1~GeV. 
At the lowest measured \pT\ the efficiency decreases by about 10\%. 
The decrease of the efficiency with centrality (up to 10\% at $|\eta|>2$) is predominantly due to the increased 
occupancy in the SCT. Residual secondaries and some fake tracks, contributing up to 5\% at the lowest 
measured \pT, are subtracted based on the HIJING sample. 
The correction due to finite momentum resolution, reaching 3\% at \pT=30 GeV, has been applied based 
on the detector response studies made with the HIJING+jet sample.

Systematic uncertainties can be attributed to several different sources. Tracking cuts and vertex 
pointing contribute up to 4\% each. The Monte-Carlo related uncertainties such as tracking efficiency, truth 
particle association, background subtraction at low \pT\ and the detector material
contribute up to 5\% depending on \pT. Many of these contributions completely or partially 
cancel when using \Rcp. The main contribution to the systematic uncertainty comes from the differences 
in the track impact parameter errors of 20-50\% between data and Monte-Carlo. It contributes up to 30\% uncertainty in the most 
central collisions. The uncertainty on \Ncoll, coming mainly from the uncertainty in the correspondence 
of the measured cross section and Glauber model cross section, contributes from 3.8\% in peripheral to 11.7\% 
in the most central \Rcp\ presented in these proceeding.

\section{Results and Summary}
\begin{figure}[t!]
\begin{center}
\includegraphics[width=0.95\textwidth]{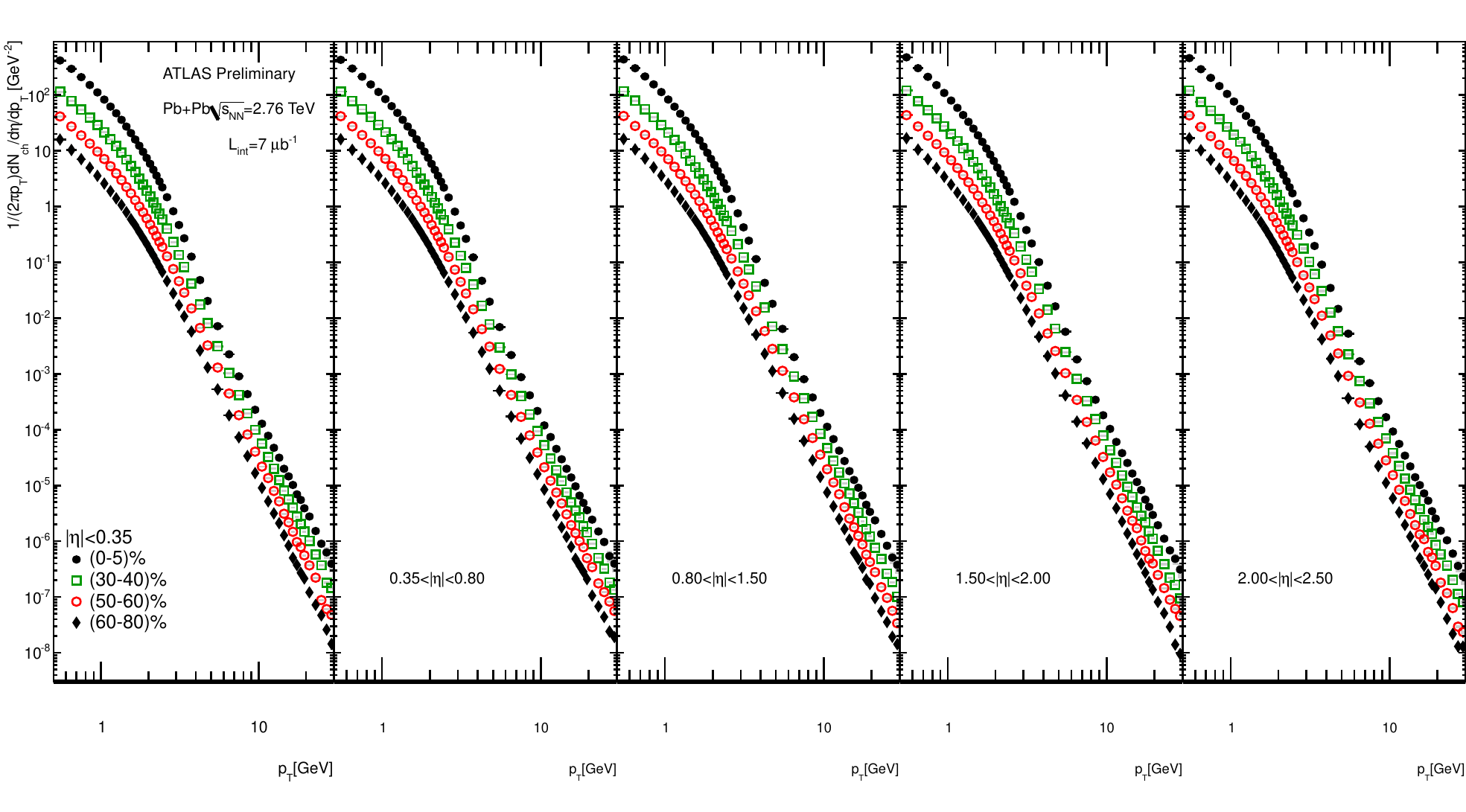} 
\caption{Fully corrected transverse momentum spectra for charged particles for four centrality 
classes and five $\eta$ ranges.
\label{fig:final_fig_spectra_all} 
}
\end{center}
\end{figure}

The corrected spectra are shown in Fig.~\ref{fig:final_fig_spectra_all}.
Invariant yields have been measured in five $\eta$ bins for different collision centrality as a function of particle \pT.
The ALICE~\cite{Aamodt:2010jd} and ATLAS central (0-5\%) yields of charged particles are consistent to within 6-7\% 
and the peripheral yields differ by more than 20\%. However, this difference can be due to differences in the definition 
of centrality classes which affect the number of \Ncoll. Including this factor the results are consistent within the 
stated uncertainties.

\begin{figure}[t!]
\begin{center}
\includegraphics[width=0.95\textwidth]{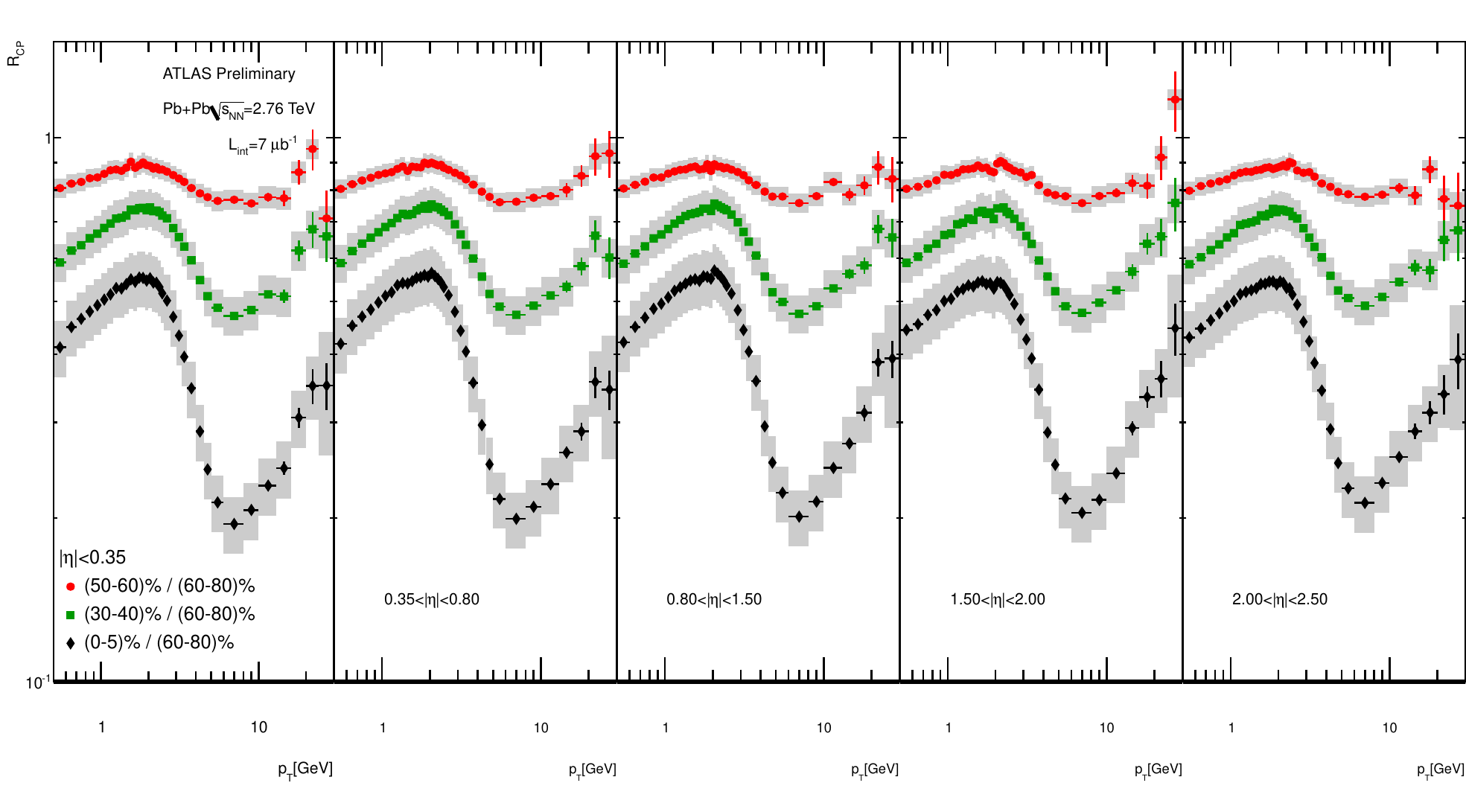} 
\vspace{-2mm}
\caption{$\Rcp$ measured in individual rapidity slices. The binning is as in Fig.~\ref{fig:final_fig_spectra_all} 
\label{fig:final_fig_rcp_all}
\vspace{-2mm}
}
\end{center}
\end{figure}

Figure~\ref{fig:final_fig_rcp_all} shows the \Rcp\ measured for different centralities. 
The spectra measured in the (60-80)\% centrality interval, shown in Fig.~\ref{fig:final_fig_spectra_all}, is used as 
the denominator in \Rcp.
The nuclear modification factor \Rcp\ drops to values comparable to the ones measured at RHIC and then 
increases up to values close to 0.5 at the upper end of \pT\ range at 30 GeV for the most central events. 
The \Rcp\ does not show any strong $\eta$-dependence.

\begin{figure}[htb]
\begin{center}
\includegraphics[width=0.46\textwidth]{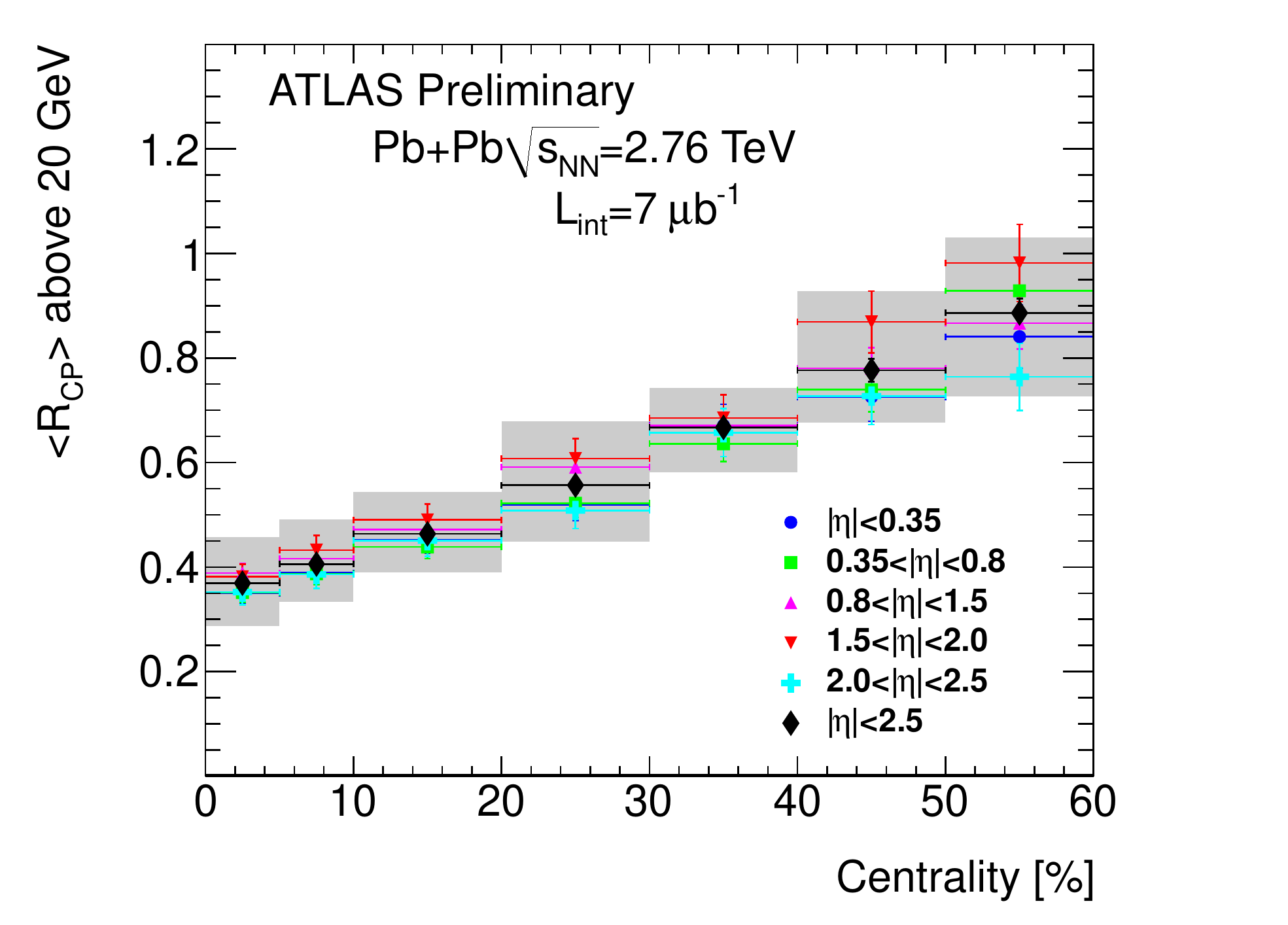} 
\vspace{-2mm}
\caption{$\Rcp$($\langle \Npart \rangle$) for spectra averaged over different $\eta$ ranges and 
above a fixed $\pT$ cut of 20 GeV as a function of centrality.  Statistical errors are shown with vertical lines and the overall systematic uncertainty at each point is shown with gray boxes.
\label{fig:rcp_above20_25_centr}
\vspace{-2mm}
}
\end{center}
\end{figure}

The \Rcp\ averaged over the entire $\eta$ range and integrated over \pT\ from 20 to 30~GeV are shown in 
Fig.~\ref{fig:rcp_above20_25_centr}. The behavior of the integrated \Rcp\ is very similar to \Rcp\ 
of the fully reconstructed jets measured by ATLAS~\cite{aaron}.

This research is supported by FP7-PEOPLE-IRG (grant 710398), Minerva Foundation (grant 7105690) and by the Israel Science Foundation (grant 710743).

\end{document}